\documentclass[11pt, a4paper]{article}
\pdfoutput=1
\usepackage{jheppub}
\usepackage{graphicx}
\usepackage{epsfig,amsmath,amsthm}
\usepackage{epstopdf}

\newcommand{\be}{\begin{equation}}
\newcommand{\ee}{\end{equation}}
\newcommand{\bea}{\begin{eqnarray}}
\newcommand{\eea}{\end{eqnarray}}
\def\bse{\begin{subequations}}
\def\ese{\end{subequations}}
 
 \newcommand{\IQ}{\mathbb{Q}}
\def\IZ{\relax\ifmmode\hbox{Z\kern-.4em Z}\else{Z\kern-.4em Z}\fi}
 \newcommand{\IT}{{\bf T}}

\newcommand{\non}{\nonumber \\}

\def\half{\frac{1}{2}} 

\def\del{{\partial}}

  \def\brho{{\bar \rho}}

  \def\eps{\epsilon}

 \def\sig{\sigma}

\def\presub{\vspace{.5cm} \noindent}

\def\bi{\begin{itemize}} \def\ei{\end{itemize}}

\def\Schw{Schwarzschild }
\def\({\left(} \def\){\right)}
\def\[{\left[} \def\]{\right]}
\def\<{\left<} \def\>{\right>}

\def\w{\omega}
\def\Or{\Omega_r} \def\Op{\Omega_\phi}
\def\sigh{\sigma_{-1/2}}

\title{Self force from equivalent periodic sources}

\author{Barak Kol\\
{\it Racah Institute of Physics, Hebrew University, Jerusalem 91904, Israel} \\
{\tt barak.kol@mail.huji.ac.il}
}

\abstract{Considering the self force and radiation due to a small body in orbit (especially aperiodic) around a black hole, this paper defines a decomposition of the source  into a sum over the shape preserving periodic motions of extended objects defined here consisting of ensembles of freely falling point particles. The stationary component is found to be within the conservative sector. Self-force computation throughout a specified trajectory reduces to solving one spatial partial differential equation (PDE) and a series of PDE's with periodic time. Regularization is alleviated due to the sources' extended nature. A regularization method is suggested inspired by the type present in electrostatics with surface or other singular sources. }

\begin{document}
\maketitle

\hspace*{\fill} \begin{minipage}{6cm}  
In memory of \\
Avraham (Avi) Schiller \\
The Racah Institute of Physics \\
1963 --- 2013 
 \end{minipage}

\section{Introduction}

The two-body problem in Einstein's gravity is important both for its natural theoretical interest and for its experimental role as an expected long-term source of gravitational waves, whose detection is the objective of a worldwide effort, see for example \cite{IFOprojects}. 

While this system can be solved numerically, analytical expansions are also available in two limits: the post-Newtonian approximation and the extreme mass ratio (EMR) limit. The latter, which is the subject of this work, has the advantage of addressing the relativistic, strong gravity physics during the late inspiral phase which is excluded by the post-Newtonian limit.

The leading correction to the geodesic motion in the probe approximation is known as the self-force, and its computation attracted considerable research especially during the last two decades, see the reviews \cite{SFEMR-rev} and references therein.  One of the central issues has been the regularization of divergences in the self-force arising from the contribution of the immediate past: the MiSaTaQuWa \cite{MiSaTaQuWa} method accomplished the regularization, at least conceptually; Barack and Ori introduced the concrete mode-sum regularization method \cite{BarackOri}; and finally Detweiler and Whiting \cite{DetweilerWhiting} introduced a useful decomposition of the field into a singular and regular parts such that the singular part does not contribute to the self-force. 

Despite the progress we identify several issues for improvement in the current method of computation. \bi

\item Length of calculation.

Presently some of the computations of the self-force require to integrate over the whole past of the body together with careful regularization. This computation needs to be repeated for each instant of time in order to achieve time evolution which accounts for the self-force.

\item Decomposition into conservative and dissipative sectors.

In \cite{HindererFlanagan} the conservative part of the self-force was defined to be the result of time-even propagation while the dissipative part  was associated with odd propagation. 
Recently another characterization of this decomposition was given for the radiation reaction force within the post-Newtonian context \cite{PNRR}. There it was proposed to define the dissipative forces to be  those which arise from the elimination of the radiation zone fields. It raises the question whether a similar characterization could be suggested in the EMR case.

\item Optimization over regularization.

Currently several variants of regularization are being used and it would be interesting to converge on an optimal one.



\ei

Let us define the goal to be the determination of the adiabatic flow due to the self-force in the space of bounded geodesic trajectories \footnote{Conveniently parametrized by the drift in action and angle variables, which in the \Schw case amounts to $(E,\, l,\, t_0,\, \phi_0)$.}. This is exactly what is required in order to describe the adiabatic inspiral phase at leading EMR order, which is the main focus of work on the gravitational self-force, but it does not include self-force corrections to scattering trajectories or plunge trajectories. 

In section \ref{sec:method} we consider quasi-periodic motion and we arrive at our main result, the suggested method of equivalent periodic sources.  In section \ref{sec:reg} we formulate the field equations throughout the trajectory and rather briefly suggest a regularization. In section \ref{sec:demonstration} we discuss present and possible future demonstrations of the method, and we conclude in section \ref{sec:sumdis} with a summary and discussion. Appendix \ref{sec:half-plane} contains an analysis of regularization in a specific and relevant case.

\section{Quasi-periodic motion and its periodic components}
\label{sec:method} 

For the purpose of presenting the method of equivalent periodic sources it suffices to consider the scalar self-force in curved space-time as follows. We consider a non-rotating (Schwarzschild) black hole of mass $M$, and a much smaller object of mass $\mu \ll M$ orbiting it in a trajectory $x^\mu = x^\mu(\tau)$. The body $\mu$ is charged under a scalar field $\Phi$ and the total action is given by \be
 S[x^\mu,\Phi] = -\frac{1}{8 \pi} \int \sqrt{-g}\, d^4x\: g^{\mu\nu}(x)\,\del_\mu \Phi\,  \del_\nu \Phi - \mu \int d\tau - q \int d\tau\, \Phi\(x^\mu(\tau)\) ~. \label{def:action} \ee
$g_{\mu\nu}$ is the black hole metric, and for concreteness we choose the \Schw gauge \bea
 ds^2 &=& - f(r)\, dt^2 + f(r)^{-1}\, dr^2 + r^2 \( d\theta^2 + \sin^2 \theta\, d\phi^2\) \non 
f(r) &:=&  1-\frac{2M}{r} ~.
\eea
Furthermore, in (\ref{def:action}) $q$ denotes the scalar charge of $\mu$ (assumed to be constant) and $d\tau=\(- g_{\mu\nu}(x)\, dx^\mu\, dx^\nu\)^{1/2}$.

Let us recall some properties of geodesic trajectories in the \Schw background. The motion is planar and the angular coordinates can be chosen such that the plane is the equatorial plane $\theta=\pi/2$. The motion is characterized by $E$ and  $l$, the energy and angular momentum per unit mass. The motion can be reduced to a radial problem and described by the following system of differential equations \bea
\( \frac{dr}{d\tau} \)^2 &=& E^2 -1 - 2\, V_{eff}(r) \non
V_{eff}(r) &:=& \frac{f}{2} \( 1 + \frac{l^2}{r^2} \) -\half = -\frac{M}{r} + \frac{l^2}{2 r^2} - \frac{M\, l^2}{r^3} \non
\frac{d\phi}{d\tau} &=& \frac{l}{r^2} \non
\frac{dt}{d\tau} &=& f^{-1}\, E \label{eom} ~.
\eea
The solution can be reduced to integrations as follows \bea
t-t_0 &=& \int_{r_0} \frac{dr}{\sqrt{E^2-1-V_{eff}(r)}} \frac{E}{f(r)} \non
\phi - \phi_0 &=& \int_{t_0} \frac{l}{r(t)^2} \frac{f(r(t))}{E}\, dt 
\eea
where the initial conditions are \bea
 r(t_0) &=& r_0 \non
 \phi(t_0) &=& \phi_0 ~. \eea

The radial motion $r=r(t)$ is periodic and hence the $r$ coordinate can be considered circular, passing twice through each value of $r$ (apart for the endpoints). The radial period is \be
 P_t =  \oint \frac{dr}{\sqrt{E^2-1-V_{eff}(r)}} \frac{E}{f(r)} ~, \qquad \oint dr \equiv 2 \int_{r_{min}}^{r_{max}} dr ~.
 \ee
During each such period the azimuthal angle $\phi$ advances by \be
P_\phi =  \oint \frac{l}{r(t)^2} \frac{f(r(t))}{E}\, dt ~, \qquad \oint dt \equiv \int_{t_0}^{t_0+P_t} dt ~. \ee
Accordingly one defines two angular velocities \bea
 \Or    &:=& \frac{2\pi}{P_t} \non
 \Op &:=& \frac{P_\phi}{P_t} \label{def:freq} ~. \eea
$\Or$ is the angular velocity associated with the radial motion and $\Op$ is the average azimuthal angular velocity. In order to stress the cyclical nature of $r$ we may define \be
 \sigma(r) := \Or\, t(r) \label{def:sigma} \ee
 which is an angle (taking values in $0 \le \sigma \le 2\pi$). From this perspective the $(r,\phi)$ space with $r$ double covered in seen as a torus $\IT^2_{\sigma\phi}$. Time evolution for fixed $E,\, l$ defines a $\phi$-invariant vector field on $\IT^2_{\sigma\phi}$. 

Quasi-periodicity means that a system contains a finite number of frequencies and their overtones (linear combinations with integer coefficients) \cite{quasiperiodic}.  Hence, by definition, the motion is quasi-periodic. It is periodic if and only if $\Or,\, \Op$ are commensurate, namely \be
 \frac{\Op}{\Or} \in \IQ ~. \ee
The aperiodic motion is known to be ergodic which implies that averages over time can be replaced by averages over appropriate ensembles.
 
\presub {\bf Phase function and frequency spectrum}. The quasi-periodic nature of the motion suggests to employ the frequency (Fourier) space. Given a Fourier frequency $\w$  the generated fields are blind to a folding of the source trajectory into a space-time with period $P=2\pi/\w$. This means that we wish to project the $t$ coordinate onto a periodic coordinate $t_\psi$ taking values in $0 \le t_\psi \le P$ and defined by \be
t_\psi = P \left\{ \frac{t}{P} \right\} \ee
where  $\{x\}$ denotes the fractional part of $x$, $0 \le \{x\} \le 1$. The periodic $t_\psi$ space-time can be visualized by identifying the $t_\psi=P$ and $t_\psi=0$ planes such that a trajectory which evolves from $t_\psi=0$ to $t_\psi=P$ re-enters at $t_\psi=0$ to continue its evolution.    

The same folded trajectory can be gotten by collecting copies of the body at locations corresponding to the instances $t= k\, P\, , ~ k \in \IZ$,
placing them all at $t_\psi=0$ and then evolving them together for one full period $0 \le t_\psi \le P$. Physically the collection can be imagined by considering the EMR binary to be moving in the dark, periodically illuminated by a stroboscope with period $P$ and recorded by long exposure camera to reveal the series of copies of the original body.

Given the Fourier frequency $\w$ the Fourier transform phase is defined over the trajectory as \be
 \psi := \omega\,  t \equiv \w\, t_\psi \label{def:psi} ~.\ee
where $t=t(\sig,\phi)$ is considered as a function over the trajectory. For ergodic motion the trajectory is dense inside $\IT^2_{\sigma\phi}$ and it is natural to seek when it extends to a continuous function $\psi(\sig,\phi)$.  After each radial period the body returns to the same radial coordinate, only to a different azimuthal angle $\phi$ and with a different phase $\psi$. In order to obtain constructive interference and hence a continuous $\psi$ the phase acquired in a full radial period should satisfy \be
 \omega P_t \equiv P_{\psi} = m\, P_\phi ~ \mbox{mod } 2\pi \label{om-cond} \ee
for some integer $m$, namely 
\be
 \omega P_t = m\, P_\phi + 2\pi\, n \label{om-cond2} \ee
  for some integer $n$. Altogether we obtain a condition for the frequencies \be
 \omega_{mn} = m\, \Op + n\, \Or   \label{freq-spec} ~, \ee
where $m,n \in \IZ$ and we used the definitions (\ref{def:freq}). Later we shall identify these frequencies to be the frequency spectrum of the motion. Its form characterizes quasi-periodic motion (in general the dimension of the frequency lattice is $\ge 2$).

For $\w=\w_{mn}$ the phase $\psi(\sig,\phi)$ can be re-expressed as  \be
\psi_{mn}(\sig,\phi) = \w_{mn} \, t(\sig) + m \( \phi -\phi(\sig)\) = \(m \frac{\Op}{\Or} + n \) \sig + m \( \phi -\phi(\sig)\) \label{def:psi-ens} \ee 
$\psi_{mn}(\sig,\phi)$ is well-defined as one goes around the $\phi$ and $\sig$ circles. More precisely the monodromies are \bea
 \phi \to \phi + 2\pi &\implies&  \psi_{mn} \to \psi_{mn} + 2\pi\, m \non
 \sig \to \sig + 2\pi &\implies& \psi_{mn} \to \psi_{mn} + 2\pi\, n ~. 
 \eea

Note that the dependence of $\psi_{mn}$ on $\phi$ in (\ref{def:psi-ens}) implies that if we were to decompose $\exp (i \psi_{mn})$ into spherical harmonics  $Y_{lm'}$ then
\be m'= m ~. \label{mprime}
\ee 
which explains our choice of notation.

\presub {\bf Equivalent periodic source}. The continuous phase function $\psi_{mn}$ (\ref{def:psi-ens}) has a simple physical interpretation. By definition all the stroboscope copies (which result from folding) have the same value of $t_\psi$ and hence the same value of $\psi$ (\ref{def:psi}). Conversely, for continuous $\psi$ the ergodic property implies that the following equiphase trajectory \be
\w_{mn} \, t_\psi = \psi_{mn} (\sig,\phi) \label{equiphase}
\ee
is the closure of the folded trajectory. This can be thought of as replacing a time average with an appropriate ensemble average which is a defining property of ergodic motion. 

Given $t_\psi$ (\ref{equiphase}) defines an equiphase surfaces $C_{mn}(t_\psi)$ which is a contour in $\IT_{\sig,\phi}$ and hence also in the $r,\phi$ plane.  An equivalent expression for it is \be
 0 = \w_{mn} \( t_\psi-t(\sig) \) + m \( \phi - \phi(\sig) \) ~. \label{equiphase2}
 \ee
 
For relatively prime $(m_0,n_0)$ a solution to (\ref{equiphase}) can be given in parametric form \bea
 \sigma &=& - m_0\, \alpha \non
 \phi   &=& n_0\, \alpha + \( \phi(\sigma) - \frac{P_\phi}{2\pi}\, \sigma \) \label{soln:shape}
 \eea
 in terms of a parameter $\alpha$. As the parameter proceeds in the range $0 \le \alpha \le 2 \pi$ the curves winds $(-m)$ times in the $\sigma$ (or $r$) direction and $n$ times in the $\phi$ direction. 
 If on the other hand $(m,n)=k\,(m_0,n_0)$ where $(m_0,n_0)$ are relatively prime and $|k|>1$ then one can still define a curve by (\ref{equiphase}) in which case it consists of $|k|$ copies equally spaced along $\IT^2_{\sigma\phi}$. 
 
 For $m_0 \neq 0$ the curve (\ref{equiphase2}) can be presented as \be
 \phi = \phi(t) - \w_{m_0 n_0}\, t      \label{equiphase3}    \ee
 which exposes a nice geometric interpretation: the curve can be generated as the orbit of a single particle as seen from a frame rotating with uniform angular velocity $\w_{m_0 n_0}$.

By construction the equiphase curve consists of an ensemble of freely falling particles, all with the same values of $E,\, l$, and its motion has the period $P_{mn}$. Its time evolution is particularly simple and beautiful: as can be seen from (\ref{equiphase2}) time translation is equivalent to a translation in either $\phi$ or $\sig$, and in particular for $m \neq 0$ time translation amounts to azimuthal rotation, thereby preserving the shape of the curve just like a soliton would, even though the individual constituents have an additional motion within the curve. 

Most interestingly due to the ergodic property the equiphase curve produces the same fields in the $\w_{mn}$ sector as the original body. Hence we refer to it as an equivalent periodic source. The density profile of the curve is an integral part of its definition and will be specified below (\ref{rho-m0n0}).

\presub {\bf Decomposition into base frequencies}. We shall now decompose the original motion into a sum of periodic motions. In doing so we shall go from the time domain to the frequency domain and back.

 The source in frequency space. The expression for the source $\rho=\rho(x^\mu)$ can be determined by comparing the source term in the action  (\ref{def:action}) with the standard source term \be
 \int \rho\, \Phi\, \sqrt{-g}\, d^4 x = q \int \Phi \, d\tau \label{def:source} \ee 
from which one deduces \be
 \rho(x^\mu) = q\, \frac{f(r)}{E}\, \delta\(r-r(t)\)\,  \frac{\delta\(\phi-\phi(t) \)}{r}\,  \delta(z)
 \ee
where $\delta(z) \equiv \delta(\theta-\pi/2)/r$ defines the equatorial plane and the notation is motivated by $z =r \sin \theta$.

Usually one defines the (Fourier) transform to frequency space\footnote{Through standard abuse of language we shall often refer to the angular velocity $\w$ as a frequency.}  through \be 
\rho(\w) = \int dt\, \rho(t)\, \exp (-i \w t) ~. \nonumber 
\ee However, for quasi-periodic quantities this definition would yield infinity for several frequencies (including $\w=0$). In order to obtain finite amplitudes one defines instead \be
 \rho(\w,x^i) := \< \rho \exp (-i \w t) \>_t \label{def:rho-om} \ee
 where for any function $f(t)$ we define its time average by \be
 \< f  \>_t :=\lim_{t_1 \to -\infty} \lim_{t_2 \to \infty}   \frac{1}{t_2-t_1} \int_{t_1}^{t_2} f(t)\, dt ~. \label{def:time-av}
\ee 
Note that for a periodic function of period $P_t$ this definition reduces to the Fourier decomposition normalized through a division by the period \be
 \< f  \>_t = \frac{1}{P_t} \int_0^{P_t} f(t)\, dt ~. \nonumber \ee

The frequency spectrum is the set of $\w$ values such that $\rho(\w) \neq 0$. A non-vanishing $\rho(\w)$ relies on the same constructive interference as the continuity of $\psi(\sig,\phi)$ and hence the frequency spectrum is given by $\w_{mn}$ of (\ref{freq-spec}).

Ensemble average. For aperiodic motion the time average in the definition (\ref{def:rho-om}) can be replaced by ensemble average with an averaged charge density \be
 \brho := \< \rho \>_t = \frac{q}{2\pi\, P_t\, r\,  (dr/d\tau)} \delta(z) \label{def:brho} \ee
 where $dr/d\tau$ depends on $r$ only and is given by (\ref{eom}). To reach this expression one replaces $\delta(\phi) \to 1/(2 \pi)$, $ \< \>_t \to P_t^{-1} \int dt$ which is justified by the ergodic property and changes the integration variable from $t$ to $r$. $\brho$ should be considered as a function of $\sig$ or equivalently as a two-valued function of $r$ whose two values happen to be identical.

Altogether we obtain 
\be 
 \rho_{mn}(r,\phi)  \equiv \rho(\w_{mn}) = \sum_{a=1}^2 \brho \, e^{-i \psi_{mn}(r,\phi)} ~, \label{rho-mn} \ee
where the sum is over the two crossings of each $r$, and $\rho_{mn}$ is a univalued function of $r$. In particular the stationary source is \be 
\rho_{00} = 2\, \brho ~. \label{rho00}
\ee

Reconstructing the source in the time domain. The Fourier modes (\ref{rho-mn}) 
contain all the information about the source and indeed can reconstruct it through the inverse transform \be
 \rho(t) = \sum_{m n} \rho_{mn}\, e^{i\, \w_{mn}\, t} ~. \label{freq-domain} \ee
 Certain partial sums have a nice interpretation. For $(m_0 n_0)$ relatively prime%
\footnote{Namely $(m_0 n_0) \neq (00)$ and there are no integers $|k|>1$ and $(mn)$ such that $(m_0 n_0) = k(mn)$. Mathematically $(m_0 n_0)$ belongs to a space that can be denoted by $\IZ \mathbb{P}^1$ and is defined by dividing $\IZ^2 \backslash (0,0)$ by the equivalence relation $(m_1 n_1) \simeq (m_2 n_2)$ if there exists an integer $k \neq 0$ such that either $(m_1 n_1) = k (m_2 n_2)$ or $k (m_1 n_1) = (m_2 n_2)$.}
 we can define the partial sum \be
 \rho_{m_0 n_0}(t) = \sum_{\begin{array}{c} k=-\infty \\ (mn) = k(m_0 n_0) \end{array}}^{\infty} \rho_{mn}\, e^{i\, \w_{mn}\, t} \label{def:periodic-comp} \ee
 which defines a periodic function of period $\left| P_{m_0 n_0} \right|$ through a standard inverse of a Fourier decomposition. $\w_{m_0 n_0}$ is the base frequency and the modes with $|k| >1$ are the overtones.  $\rho_{m_0 n_0}$ simplifies to \be
 \rho_{m_0 n_0}(t) = 2\pi\, \brho\,  \delta\(\w_{m_0 n_0} t - \psi_{m_0 n_0} \) \label{rho-m0n0}
 \ee
 by using (\ref{rho-mn}) and the identity $\sum_k  \exp (i\, k\, \theta) = 2 \pi \, \delta (\theta)$. This expression describes the motion of the curve $C_{m_0 n_0}$ together with its $2 \pi\, \brho$ density profile which is equivalent to a uniform density in $\IT_{\sig,\phi}$, restricted to the curve. 
 
Since the stationary $(00)$ mode belongs to every partial sum $\rho_{m_0 n_0}(t)$ we define \be
  \rho'_{m_0 n_0}(t)= \rho_{m_0 n_0}(t) - \rho_{00} \label{def:rho'}
 \ee
 to avoid double counting. These definitions allow us to return to the time domain and define the final decomposition of the source to be  \be
 \rho(t) = \rho_{00} + \sum_{m_0,n_0} \rho'_{m_0 n_0}(t) \label{def:decomp} \ee
  where the sum is over $(m_0,n_0)$ which are relatively prime. It expresses $\rho(t)$ in terms of the time-independent component and several periodic components associated each with an equivalent source. 
  
This decomposition has a hybrid character of both time and frequency domains: it contains a sum over frequencies but all the components are in the time domain. We can argue for preferring the partial inverse transform (\ref{def:periodic-comp}) over the full frequency domain (\ref{freq-domain}) by noting that as the equivalent source passes through a point in space it creates a delta function source in time, which is easier to analyze in the time domain (in principle, frequency space is useful for functions with few significant Fourier coefficients unlike the delta function).

\begin{figure}[t]
\centering
\includegraphics[width=17cm]{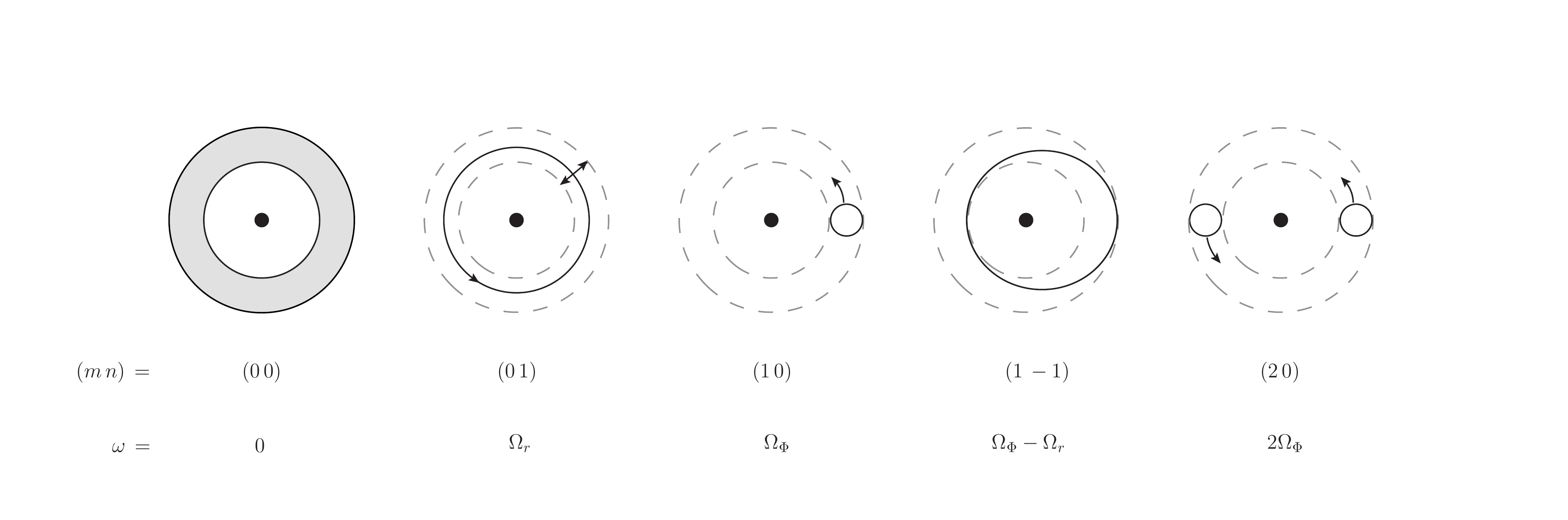}
\caption{Examples of equivalent periodic sources for several frequencies.The small solid disc at the center represents the black hole and the dashed grey lines represent $r_{min},\, r_{max}$.}
\label{fig:examples}
\end{figure}

\presub {\bf Examples}. We proceed to illustrate the equivalent periodic source for certain specific frequencies, see fig. \ref{fig:examples}. 
 
We start with the case $\omega_{00} \equiv 0$. In this case the equivalent source is a stationary 2d charge density. It emits no radiation and therefore the associated self-force is conservative. 

The next case we consider is $\omega_{01} \equiv \Or$. The equivalent source are uniformly distributed $r=const$ curves (\ref{equiphase2}) and the motion consists of radial pulsations. The emitted radiation is suppressed due to its $m=0$ nature and the rather limited amplitude. 

For $\omega_{10} \equiv \Op$ the equivalent source is a closed curve which spans the range from $r_{min}$ to $r_{max}$, but does not wind around $M$. In fact, this is the only equivalent source which does not wind $\phi$. Its density profile is set by $\brho$ (\ref{def:brho}) As time evolves it orbits $M$ without changing its shape.   For a given $r$ value the $\phi$ separation between the two crossings of the curve determines the loss of amplitude due to interference -- see (\ref{rho-mn})

For $\omega_{1,-1} \equiv \Op-\Or$ the orbit winds once both in the $r$ direction and in the $\phi$. In the large $r$ post-Newtonian limit the equivalent source is very  close to the actual trajectory. It evolves in time due to precession of the periastron, whose small frequency is indeed $\Op-\Or$. This low frequency suppresses the radiation from this mode.

Finally $\omega_{20} \equiv 2 \Op$ is an example for non relatively prime $(mn)$ and it contains a pair of the $(10)$ sources separated in the azimuthal direction by half a circle.

\section{Field equations, regularization and self-force}
\label{sec:reg}

{\bf Field equations}. Having decomposed the source into periodic components (\ref{def:decomp}) we proceed to obtain the field equations. Since for bounded geodesic motion the source is quasi-periodic, so are the generated fields, possibly after transients have decayed. The analogous decomposition of the field reads \be 
  \Phi(t) = \Phi_{00} + \sum_{m_0,n_0} \Phi'_{m_0 n_0}(t) \qquad  \Phi'_{m_0 n_0}(t)= \Phi_{m_0 n_0}(t) - \Phi_{00}   \label{def:Phi-decomp}
\ee
 The $\Phi$ equation of motion derived from the action (\ref{def:action}) together with (\ref{def:source}) decomposes into  \bea
  \triangle \Phi_{00}  &=& 4 \pi\, \rho_{00} \non
 \square \Phi'_{m_0 n_0}  &=& 4 \pi\, \rho'_{m_0 n_0} \label{Phi-eom} \eea
 where\footnote{%
 In general $\Phi$ can also have a homogenous wave component, which is assumed here to be absent due to a boundary condition of no incoming waves. Such a component would have been analogous to transients in the forced oscillator.}
 \footnote{%
 The operator $\triangle$ is not the Laplacian for some curved spatial geometry. However,  we can express is as $\triangle=|g_{tt}| \, \tilde{\triangle}$ 
 where $\tilde{\triangle} $ \emph{is} the Laplacian for the spatial metric defined through the standard post-Newtonian decomposition \cite{CLEFT-caged,NRG} namely $\gamma_{ij}:=|g_{tt}|\, g_{ij}$.}
 \bea
 \triangle \Phi &:=& \frac{1}{\sqrt{-g}} \del_i\, \sqrt{-g}\, g^{ij}\, \del_j\, \Phi \non
 \square &=& - f^{-1}\, \del_t^2 + \triangle  ~.
 \eea

The field equation simplifies in the common case when the source has a helical symmetry \be
\Phi(\phi, t) = \Phi (m\, \phi - \w\, t) \label{def:helical} ~. \ee
 This occurs for circular motion where $m=1,\, \w=\Omega$ as well as for all the components of an aperiodic motion where $m=m_0,\, \w=\w_{m0 n0}$.\footnote{The radial pulsations component, $m_0=0$, is not helical in the usual sense but it does satisfy (\ref{def:helical}).} The helical symmetry implies the time periodic field equation are  3d rather than 4d, while the stationary sector is 2d rather than 3d.

{\bf Boundary conditions}. In the stationary sector the boundary conditions are asymptotic regularity \be
 \Phi_{00} \sim Q \cdot \frac{1}{r} + 0 \cdot 1 \mbox{ for } r \to \infty \label{bc-hor1}
\ee
for some constant Q, and a Gauss constraint at the horizon \be
\oint d\Omega\, \left. \del_r \Phi_{00} \right|_{r=r_0} = 0 \label{bc-hor2}
\ee
which guarantees that the black hole has zero scalar charge.

In the periodic sectors the boundary conditions are no-incoming radiation both asymptotically and from the horizon, and this is expressed by \bea
 \Phi &\sim& \frac{\Phi^{out}_\infty(\theta,\phi,t-r)}{r} + 0 \cdot \frac{\Phi^{in}_\infty(\theta,\phi,t+r)}{r} \mbox{ for } r \to \infty \non
 \Phi &\sim& \Phi^{out}_h(\theta,\phi,t+r^*)  + 0 \cdot \Phi^{in}_h(\theta,\phi,t-r^*) \mbox{ for } r \to r_0 \label{Phi-bc} \label{bc-asymp} \eea
where $\Phi$ abbreviates $\Phi'_{m_0 n_0}$, the outgoing waves are periodic with period $P_{m_0 n_0}$ and $r^*$ is the usual radial tortoise coordinate defined by $dr^*:=dr/f(r)$.

\presub {\bf Regularization}. The specification of the equation of motion (\ref{Phi-eom}) together with the boundary conditions (\ref{bc-hor1}-\ref {Phi-bc}) would have completely defined the solution, except for the need to handle the singular nature of the source $\rho$. This is essentially the well-known problem of self-force regularization. It is natural to address regularization separately in each periodic sector, where the equivalent source, rather than the point particle, is the correct singular locus.

Minimal regularization. We suggest a regularization inspired by the one used in electrostatics with singular sources. The field (in each sector) is decomposed into a singular and a regular part \be
\Phi = \Phi_S + \Phi_R ~. \label{RS} \ee
 The singular part $\Phi_S$ must be an analytic function chosen to solve the equation up to a regular (R) remainder, namely $\Phi_S$ should be such that  $\rho_R$ defined by
 \be
  4 \pi\, \rho_R := 4 \pi \rho - \square \Phi_S  \ee
is regular. Clearly $\Phi_S$ is unique only up to adding a regular function. 

We suggest to choose $\Phi_S$ to be as simple as possible as long as it has the right local structure and it satisfies any known analytic behavior at the boundaries.  For instance, for aperiodic motion with a 1d equivalent source $\Phi_S$ behaves in the vicinity of the source as \be
 \Phi_S(x) \sim 2\, \lambda\, \log \Gamma(x) \ee
where $\Gamma$ is the  square of the (geodesic) distance from $x$ to the world-tube swept by the $(m_0 n_0)$ curve and $\lambda$ is the line charge density close to $x$. We refer to this regularization as ``minimal''. It differs from the popular Detweiler-Whiting decomposition \cite{DetweilerWhiting} as here $\Phi_S$ is allowed to contribute to the self-force.
More generally all the possible singular terms in $\Phi_S$ can be determined by dimensional analysis. This is discussed and demonstrated in a particular and relevant case in appendix \ref{sec:half-plane}. A full method to obtain $\Phi_S$ remains to be developed.

Numerical evaluation of $\Phi_R$. Next one solves (numerically) for $\Phi_R$ \be
\square \Phi_R = 4 \pi\, \rho_R ~.
\ee
In the stationary case the equations are to be solved locally, in the spirit of relaxation \cite{relaxation}, rather than through integrating over Green response functions. We are interested in the value of $\Phi$ not only at a specific point, but over the whole trajectory, which will be useful for computing the self-force throughout the motion. In this case a local method of solution appears more efficient. The algorithm for solving the time-periodic equations may need to be developed.

\presub Having found the equivalent source (\ref{rho-mn}) and the resulting field from eq. (\ref{Phi-eom}) we can proceed to determine both the outgoing radiation and the self-force.

{\bf Outgoing radiation}. While discussing the boundary conditions in (\ref{Phi-bc}) we defined \be 
\Phi_\infty^{out}(\theta,\phi,t-r),\, \Phi_h^{out}(\theta,\phi,t+r^*)
\ee
 which  characterize the outgoing radiation both asymptotically and into the horizon. Given a solution for the field $\Phi$ one can determine $\Phi_\infty^{out}(\theta,\phi,t-r),\, \Phi_h^{out}(\theta,\phi,t+r^*)$, and from them the conserved quantities carried away by the radiation, namely the angular momentum, linear momentum and especially the energy. This can be done through evaluation of $T_{\mu\nu}$ the energy-momentum tensor associated with the field, or equivalently using expressions in the literature such as \cite{RossMultipoles}.

{\bf Self-force}. The force acting on the body at any given instant due to the field, namely the self-force, can be obtained by varying the action (\ref{def:action}) with respect to $x^\mu$. One must include both the finite part of the analytic part $\Phi_S$ as well as the numerical regular part $\Phi_R$.

As an example for a dissipated quantity let us consider the energy. The expression for the momentary self-force can be averaged over time to yield the average loss of power. The ergodic property allows to replace the time average with the ensemble average, which is also equivalent to the power dissipated by the equivalent periodic source.

The balance of energy between the outgoing radiation and the dissipation can be understood now as follows. Both quantities can be computed through the equivalent periodic source, where there is a single base frequency and hence the power must be balanced for each frequency separately.

{\bf The stationary sector}. The only sector where there is no outgoing radiation  is $\w=0$. Therefore there can be no dissipation and we conclude that \be
 \w=0  \Longrightarrow  \mbox{conservative }
 \ee
The surface charge density is given by (\ref{rho00}). The non-stationary sources $\rho'$ (\ref{def:rho'})  are wholly responsible for the dissipative effects.

\section{Demonstration} 
 \label{sec:demonstration}

The first arXiv version of this paper included various suggestions for demonstrating the method. Luckily, it turns out that a rather strong demonstration already exists.  In the last year Hod suggested an ``analytic toy model'' to compute the shift in the innermost circular orbit (ISCO) \cite{Hod}.\footnote{I am thankful to L. Barack for pointing out this work to me.} %
Within this model a body in circular orbit is replaced by a ring. Then analytic results for the gravitational self-force obtained in \cite{Will} are applied yielding a relative shift of the ISCO frequency by $c_A\, \mu/M$ with the analytic constant $c_A = 1 + 29/(81 \sqrt{2}) \simeq 1.253$. This result is very close to the numerical value $c_N \simeq 1.251$ see \cite{ABDS} and additional references within \cite{Hod}. The author of \cite{Hod} notes (see summary section) that the original motivation was to provide a simple qualitative explanation for the sign of the shift, and admittedly there was no reason to expect the astonishingly close quantitative agreement. 

From the point of view of the current paper the ring is precisely the equivalent source in the stationary sector, which contributes to the conservative self-force and hence to the shift in frequency. Presumably the contribution of the other sector (sourced by $\rho'$) is much smaller. Hence the apparently coincidental high accuracy agreement is mostly understood and serves as a demonstration of the idea of equivalent sources.

We now turn from a past work which turned into a demonstration, into a case by case analysis of several possible future demonstrations.

Keplerian motion. In this case the motion is periodic and hence the decomposition (\ref{def:decomp}) includes only two sectors: the stationary sector, which is purely conservative, and the $\w \neq 0$ sector sourced by $\rho'$ (\ref{def:rho'}). For $\w=0$ one needs to solve a spatial equation in 3d (\ref{Phi-eom}) with a source line. For $\w \neq 0$ the field equation is 4d (with periodic time), the source being a world-line.

Circular motion. In addition to being periodic this case enjoys the helical symmetry (\ref{def:helical}) and hence the field equations can be further reduced by symmetry such that for $\w=0$ one solves a 2d equation in the $(r,\theta)$ plane with a point source and for $\w \neq 0$ one solves a 3d spatial equation with a point-source. In this case the self-force is also particularly simple. Being contained in the equatorial plane and being orthogonal to the 4-velocity the self-force has only two independent components: a conservative component in the $r$ direction and a dissipative component in the $t,\phi$ plane. These two components are time-independent thanks to the helical symmetry. In the current formulation this constant self-force separates into $\w=0$ and $\w \neq 0$ components, the former being purely radial.

General aperiodic motion. The $\w=0$ sector is $\phi$ independent and hence requires solving a 2d equation in the $(r,\theta)$ plane with a segment source. It is usually an important part of the conservative source. Next one computes the $\w=\Op$ sector which is usually an important contributor to the dissipative sector, being the only equivalent source which does not wind the $\phi$ direction.  The $\w=\Op$ sector requires solving a 3d equation. In principle one should compute an infinite seres of $(m_0 n_0)$ periodic sources. However, their contribution to the outgoing radiation and self-force is expected to decrease and so for any goal accuracy it should suffice to compute a finite number of them. In order to judiciously choose the sectors to be computed, it could be interesting to seek estimates for the radiation and/or self-force contributions of the various sectors.

More general cases involving motion around a rotating Kerr black hole, or types of interactions other than scalar are discussed in the next section.

\section{Summary and discussion}
\label{sec:sumdis}

The main result of this paper is the definition of equivalent periodic sources and an associated decomposition of the source into periodic components, which are especially useful for bound motion with incommensurate base frequencies (and hence ergodic).  These sources were explained to be an ensemble average, which for ergodic motion is equivalent to a time average. For concreteness we analyzed a body charged under a scalar field in bound motion around a \Schw black hole. In this case for each frequency $\w_{m_0 n_0}$ (\ref{freq-spec}) the body can be replaced by an equivalent periodic source (\ref{rho-m0n0}) supported on the curve $C_{m_0 n_0}(t)$ defined  by either one of the expressions (\ref{equiphase}-\ref{equiphase3}). The curve winds $(-m)$ times in the radial direction and $n$ time in the azimuthal direction. As time evolves the shape remains essentially unchanged\footnote{See paragraph after (\ref{equiphase3}).} and the motion has the period associated with $\w_{m_0 n_0}$. 

The source decomposition is given in (\ref{def:decomp}) together with the definitions (\ref{def:periodic-comp},\ref{def:rho'}). It is a decomposition of the source into an infinite number of periodic extended sources whose periods are incommensurate. It can be thought to be a hybrid of the frequency and time domains. The source decomposition applies to the electromagnetic and gravitational cases as well, and a similar theory holds for bound motion around other black holes.

Some additional results. The back-reaction field is analogously decomposed (\ref{def:Phi-decomp}) together with the formulation of the equations of motion (\ref{Phi-eom}) and boundary conditions obeyed by the components (\ref{bc-hor1}-\ref{bc-asymp}). The field components $\Phi_{m_0 n_0}$ are solved for throughout space, and in particular throughout the trajectory, thus enabling to compute both the outgoing radiation and the adiabatic drift of the trajectory in trajectory space due to the self-force.  A brief definition of a regularization method is given in (\ref{RS}) inspired by the electrostatics of singular sources. Of course the method of regularization is an independent ingredient in the formulation, as any (correct) regularization would yield correct results, though the computational cost may differ.

Let us revisit the issues from the introduction. The first issue was length of calculation. In the suggested method the field is solved throughout space thereby providing the self force throughout the trajectory. In addition an infinite series of frequencies is summed into a periodic function and solved at once. This promises to be more efficient since the frequency domain resembles the Fourier decomposition of a delta function in time and as such is not decreasing fast. The second issue was a decomposition into conservative and dissipative parts. Here we were able to associate the stationary sector with the conservative part, but the issue may not be fully resolved. The final issue was that of optimizing over regularization. Here we suggested a minimal regularization, but it would require further development. Altogether the main result evolved from these issues, but not in a direct manner.

Extant literature. The self force for an eccentric \Schw trajectory was computed in \cite{WarburtonBarack} using the frequency domain and mode sum regularization. Their eq. (17) appears to be the Fourier transform of the equivalent source (\ref{rho-m0n0}). \cite{VWDCH} treated the same problem, though with different methods: the used the effective source method together with ``a reasonably accurate approximation to the Detweiler-Whiting singular field'' and they introduced the notion of self force loops. 

 \presub Let us proceed to discuss some points.

{\bf The gravitational self-force}. While we chose to present the ideas in the simple setting of the scalar self-force we are mostly interested in the gravitational case, and possibly in the electromagnetic. Here we would like to outline some of the changes which occur in the generalization to these cases. For brevity we shall denote by $s=0,1,\,2$ the scalar, electromagnetic and gravitational cases  ($s$ would be the spin of the field in the quantized theory).

The equivalent periodic source is the same for all cases. Naturally the source $\rho_{m_0 n_0}$ will be replaced by $j^\mu_{m_0 n_0}$ for $s=1$ and $T^{\mu\nu}_{m_0 n_0}$ for $s=2$, and both are constrained by conservation of charge/ stress-energy. Outgoing radiation contains now two polarizations (for both $s=1,2$) and is bounded from below by $l \ge s$.  The stationary sector includes in addition to these also the $l<s$ black hole modes. For $s=1$ these reflect contributions to the black hole electric and magnetic charges, and for $s=2$ these are the mass, momentum, center of mass and angular momentum.

\presub {\bf Motion around a more general black hole}.  Let us outline the generalization from \Schw to other black holes, the rotating Kerr black hole being the main application. For Kerr, motion is possible in the $\theta$ direction in addition to $r,\phi$, yet it remains integrable. Accordingly, there are generically 3 base frequencies and the distribution is not limited to a surface anymore. The equiphase condition (\ref{equiphase}) defines now the equivalent source to be a surface. Interestingly, this further alleviates the regularization issue, possibly obviating it altogether in the stationary sector. 

\presub {\bf Higher EMR orders}. It would be interesting to obtain higher EMR orders. The form of the equivalent source could receive corrections.

\presub {\bf Region of usefulness}. The equivalent source should be useful quite generally for any self-force computation (for an aperiodic trajectory) at leading EMR order. The non-relativistic limit should probably be avoided since it develops a second length scale (that of the wavelength) and hence is problematic for numerical evaluation. In addition, this case has a theory of its own (\cite{GalleyTiglio,PNRR}).

The case of nearly incommensurate frequencies is interesting. Nearly incommensurate means that the ratio of frequencies is a rational number, yet with a large denominator. Indeed given any non-zero numerical precision, this is the only meaning of being incommensurate. In this case the motion is periodic, but it wounds around for many times before it closes. In this case it would be a good approximation to replace the dense world-lines by a smooth surface distribution.  

\presub {\bf Conclusion}. The main result of this paper is to introduce the notion of an equivalent periodic source and to use it to define a decomposition of  the source. The provided proofs guarantee its correctness and I believe it to be novel. The correctness is independently strengthened by the demonstration in \cite{Hod} as explained in section \ref{sec:demonstration}, and it also demonstrates its utility to replace certain numerical results by analytic derivations.  

Its interest is presumably self-evident as is suggests a new way of thinking about the problem. It offers insight into the form of the source in frequency space (\ref{rho-mn}) and for the relative strength of the sectors, see section \ref{sec:demonstration}, next to last paragraph. The method offers a distinct computational formulation, a certain hybrid of the time and frequency domains. Finally, it has a potential for economizing computation and reaching new computational results since it nicely sums over any base frequency and its overtones.

\subsection*{Acknowledgments}

It is a pleasure to thank Amos Ori for discussions and comments on this work and more generally for discussions of self force over the years; Leor Barack for a discussion on the state of the art in EMR and for pointing out \cite{Hod} to me; Ofek Birhnholtz and Shahar Hadar for comments on a presentation of this work; and finally the organizers and participants of the 16th Capra  meeting (Dublin, 15-19 July 2013) where this work was presented for their challenging questions.

This research was supported by the Israel Science Foundation grant no. 812/11. 

\appendix

\section{Regularization of the potential due to a half-plane surface density}
\label{sec:half-plane}

Let us consider the following surface density over a half-plane \be
 \rho (x,y,z) = \frac{\sigma_{-1/2}}{\sqrt{x}} \delta(z) \qquad x \ge 0 ~. \label{def:sigma-half} \ee
 in flat space.
 This distribution is a local model for the equivalent average density for aperiodic motion (\ref{def:brho})  around $r_{min}$ or $r_{max}$ where $x$ in (\ref{def:sigma-half}) models the radial direction and the $\sqrt{x}$ factor originates in the behavior of $dr/d\tau$ near a radial turning point.

{\bf Existence}. We shall first discuss the existence of $\Phi_S$ and then its actual expression. In general, regularization instructs us to extract a finite answer from a divergent quantity by performing a local (Lorentz) expansion of the quantity around the point of interest and keeping only the finite part. We shall posit a working assumption that a regularization is possible if the force is finite. 

Generalities -  dimensional analysis. To examine whether the force is finite it is enough perform a local analysis, integrating over nearby force contributions. Dimensional analysis provides a necessary condition for the existence of divergences. Denote the coefficient of the term which is to be tested by $C$. Its dimension can be written as \be
[C] = \frac{Q}{L^2}\, L^a \ee
where $Q, L$ are the charge and length dimensions and $a$ is some constant, which is so defined since the dimensions of force are $Q/L^2$. A necessary condition for a divergence is \be
 a \ge 0 ~.\ee
Indeed by dimensional analysis a divergence would have the form $F \sim C/\eps^a$, where $F$ is the force, and $\eps$ is a short distance cut-off.

We note that nearby the singular source the term $\w^2 \Phi$ is negligible relative to  $\triangle  \Phi$ (which scales like $\Phi/\eps^2$) and is hence irrelevant.

Application to the case at hand. In the case (\ref{def:sigma-half}) $C$ is $\sigh$ and $a=+1/2$ which implies that a divergence is possible. Therefore we must proceed to evaluate the force locally at $x=\eps$. The only component of $F$ which can possibly diverge is $F_x$. We have \bea
F_x(\eps) &=& -\int dq \frac{x-\eps}{r^3} = - \int_0^\infty \frac{\sigh}{\sqrt{x}}\, dx\, \int_{-\infty}^\infty dy\, \frac{x-\eps}{\((x-\eps^2)+y^2\)^{3/2}} \propto \non
 &\propto& -\int \frac{\sigh}{\sqrt{x}}\,  dx \frac{1}{x-\eps} = - \frac{\sigh}{\sqrt{\eps}}\, \int_0^\infty \frac{d\xi}{\sqrt{\xi}(\xi-1)} \eea
 where the third equality can be deduced on dimensional grounds and $\xi:=x/\eps$. 
 
 As expected a divergence of the form $\sigh/\sqrt{\eps}$ is possible and one must proceed to evaluate the dimensionless integral which follows it. The integrand is divergent around $\xi=1$, however this would happen for any surface density,  not necessarily having the $q/\sqrt{x}$ profile, and can be regularized by the principal value (PV) prescription. One finds \bea
  \mbox{PV } \int_0^\infty \frac{d\xi}{\sqrt{\xi}(\xi-1)} &=& \[ \log \frac{\sqrt{\xi} -1}{\sqrt{\xi} +1} \]^\infty_{1+\delta} +  \[ \log \frac{1-\sqrt{\xi} }{1+\sqrt{\xi}} \]^{1-\delta}_0 = \non
  &=& 0 -\log \frac{\delta}{4} +\log \frac{\delta}{4} - 0 = 0 ~. \eea 
Therefore the force is finite in the presence of the charge distribution (\ref{def:sigma-half}).

 {\bf Expression for $\Phi_S$}. In the case (\ref{def:sigma-half}) complex analysis can be used to produce an analytic singular solution which is actually a full solution. Since $\rho$ is independent of $y$ it suffices to solve the problem in $(x,z)$ plane. The equation (\ref{Phi-eom}) becomes \be
  \[ \del^2_x + \del^2_z \] \Phi = 4 \pi\, \rho ~, \ee
  and the discontinuity condition is \be
 [\del_z \Phi ]_{z=0} = 2 \pi\, \frac{\sigh}{\sqrt{x}} ~.\ee
  Defining a complex coordinate \be
  w=x+i\, z \ee
 \be
  \Phi = 2 \pi\, \sigh\, \Re \sqrt{-w}  \ee
is a harmonic function satisfying the discontinuity condition and hence a solution.



\end{document}